\documentclass[preprint,12pt,a4paper]{elsarticle}

\usepackage[mathlines]{lineno}
\usepackage{amssymb}
\usepackage{amsmath}
\usepackage{hyperref}
\usepackage{ifthen}
\usepackage{mciteplus}
\newboolean{articletitles}
\setboolean{articletitles}{true}

\journal{Nuclear Instruments and Methods A}

\newcommand{\ud}{\ensuremath{\mathrm{d}}}
\newcommand{\domega}{\ensuremath{\delta_\omega}}
\newcommand{\jpsi}{\ensuremath{J/\!\psi}}
\newcommand{\mumu}{\ensuremath{\mu^+\mu^-}}
\newcommand{\Dz}{\ensuremath{D^0}}
\newcommand{\DzToKPi}{\mbox{\ensuremath{\Dz\to K^{-} \pi^+}}}
\newcommand{\JpsiToMuMu}{\mbox{\ensuremath{\jpsi\to\mumu}}}
\newcommand{\xvtx}{\ensuremath{\xi}}
\newcommand{\xtrk}{\ensuremath{x}}
\newcommand{\Cvtx}{\ensuremath{C_\xvtx}}
\newcommand{\GB}{G^{B}}
\begin{document}

\begin{frontmatter}



  \title{{\bf Application of vertex and mass constraints in track-based
    alignment}}
  
  \address[nikhef]{Nikhef National Institute for Subatomic Physics, Amsterdam, The Netherlands}
\address[heidelberg]{Physikalisches Institut, Ruprecht-Karls-Universit\"at Heidelberg, Heidelberg, Germany}
\address[syracuse]{Syracuse University, Syracuse, NY, United States}
\address[glasgow]{School of Physics and Astronomy,University of Glasgow, Glasgow, United Kingdom}
\address[manchester]{School of Physics and Astronomy,University of Manchester, Manchester, United Kingdom}
\address[cern]{European Organization for Nucleaer Research (CERN), Geneva, Switzerland}
\address[zurich]{Physik-Institut, Universit\"at Z\"urich, Z\"urich, Switzerland}
\address[epfl]{Ecole Polytechnique F\'ed\'erale de Lausanne (EPFL), Lausanne, Switzerland}
\address[liverpool]{Oliver Lodge Laboratory, University of Liverpool, Liverpool, United Kingdom}
\address[mpiheidelberg]{Max-Planck-Institut f\"ur Kernphysik (MPIK), Heidelberg, Germany}
\address[edinburgh]{School of Physics and Astronomy,University of Edinburgh, Edinburgh, United Kingdom}
\address[ferrara]{Sezione INFN di Ferrara, Ferrara, Italy}
\cortext[cor1]{Corresponding author}
\author[nikhef]{J.~Amoraal}
\author[heidelberg]{J.~Blouw}
\author[syracuse]{S.~Blusk}
\author[glasgow,manchester]{S.~Borghi}
\author[cern]{M.~Cattaneo}
\author[zurich]{N.~Chiapolini}
\author[epfl]{G.~Conti} 
\author[heidelberg]{M.~Deissenroth}
\author[epfl]{F.~Dupertuis}
\author[nikhef]{R.~van~der~Eijk}
\author[epfl]{V.~Fave}
\author[cern]{M.~Gersabeck}
\author[epfl]{A.~Hicheur}
\author[nikhef]{W.~Hulsbergen\corref{cor1}}
\ead{wouter.hulsbergen@nikhef.nl}
\author[liverpool]{D.~Hutchcroft}
\author[nikhef]{A.~Kozlinskiy}
\author[nikhef]{R.W.~Lambert}
\author[mpiheidelberg]{F.~Maciuc}
\author[epfl]{R.~M\"arki\corref{cor1}}
\ead{raphael.marki@epfl.ch}
\author[nikhef]{M.~Martinelli}
\author[nikhef]{M.~Merk}
\author[edinburgh]{M.~Needham}
\author[epfl]{L.~Nicolas}
\author[nikhef,cern]{J.~Palacios}
\author[manchester]{C.~Parkes}
\author[nikhef]{A.~Pellegrino}
\author[ferrara]{S.~Pozzi}
\author[nikhef]{G.~Raven}
\author[glasgow,manchester]{E.~Rodrigues}
\author[zurich]{C.~Salzmann}
\author[nikhef]{M.~Schiller}
\author[epfl]{O.~Schneider\corref{cor1}}
\ead{olivier.schneider@epfl.ch}
\author[nikhef]{E.~Simioni}
\author[zurich]{O.~Steinkamp}
\author[heidelberg]{J.~van~Tilburg}
\author[nikhef]{N.~Tuning}
\author[heidelberg]{U.~Uwer}
\author[ferrara]{S.~Vecchi}
\author[glasgow]{S.~Viret}

  \begin{abstract}
    The software alignment of planar tracking detectors using samples of
    charged particle trajectories may lead to global detector
    distortions that affect vertex and momentum resolution. We present
    an alignment procedure that constrains such distortions by making
    use of samples of decay vertices reconstructed from two or more
    trajectories and putting constraints on their invariant mass. We
    illustrate the method by using a sample of invariant-mass
    constrained vertices from \mbox{\DzToKPi} decays to remove a
    curvature bias in the LHCb spectrometer.
  \end{abstract}
  
  
\end{frontmatter}



\section{Introduction}
\label{sec:Introduction}

The calibration of the position and orientation of tracking detectors
in high energy physics experiments is called \emph{alignment}. The
input for alignment comes from two sources, namely survey information
collected during assembly or after installation, and hit residuals of
reconstructed charged particle trajectories (tracks). With track-based
algorithms an alignment accuracy can be reached that well exceeds the
single-hit resolution.

The track-based alignment algorithms considered here optimize the
total fit quality of a sample of tracks, for example the total track
fit $\chi^2$, with respect to a set of numbers that parametrize the
detector geometry. The parameters are usually chosen to be the positions
and orientations of individual detector elements. We denote the set of
alignment parameters with a generic symbol $a$. The condition that the
total $\chi^2$ be minimal with respect to $a$ can then be written as
\begin{linenomath}
\begin{equation}
\frac{\ud}{\ud a} \sum_\text{tracks $i$} \chi_i^2 \; = \; 0 \; ,
\end{equation}
\end{linenomath}
where the sum runs over all tracks in the calibration sample.

\begin{figure}[hbt]
  \centerline{
  \includegraphics[width=.6\columnwidth]{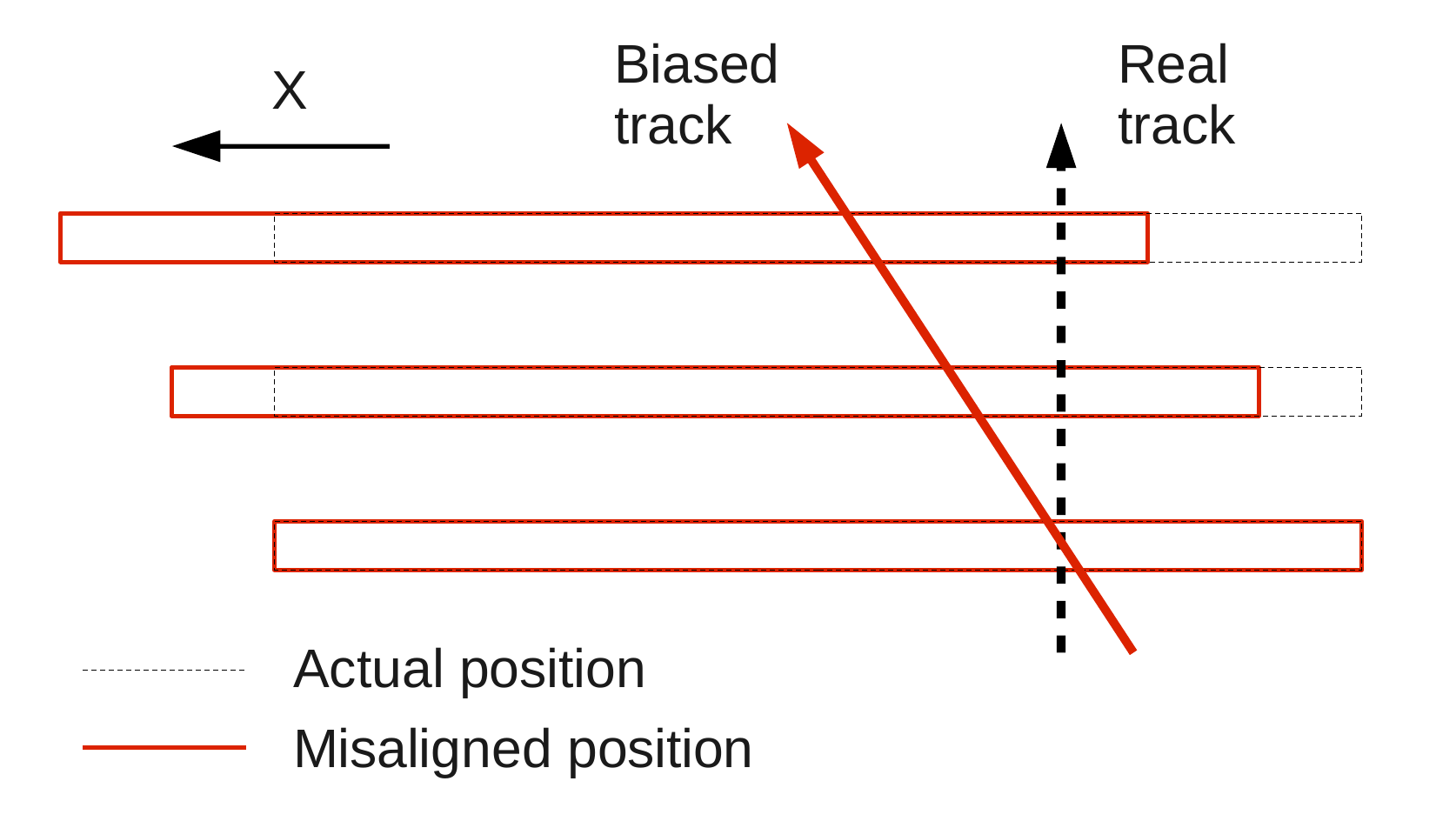}
  }
  \caption{Illustration of a shearing in a planar detector.}
  \label{fig:shearing}
\end{figure}

A common problem in the application of track-based alignment
algorithms is related to so-called weak modes. These denote alignment
degrees of freedom to which the total track $\chi^2$ is mostly or
completely insensitive. A global translation or rotation of all
detector elements with respect to a common point (i.e.\ a coordinate
transformation) can be thought of as a perfect weak mode. Less trivial weak modes are
related to global distortions and depend on the detector geometry. In
parallel plane detectors, shearings, such as that depicted in
Fig.~\ref{fig:shearing}, are also examples of weak modes.

One reason weak modes are a concern in alignment is that they lead to
poor convergence. An approach to treat weak modes is to impose
additional constraints. For example, information on alignment
parameters obtained from detector survey can be used in the
minimization procedure by including additional terms in the
$\chi^2$,
\begin{linenomath}
\begin{equation}
  \chi^2_\text{survey} \; = \; \sum_k \left( \frac{ a_k -
      a_{k,\text{survey}}}{\sigma_{k,\text{survey}}} \right)^2
  \label{equ:surveychisq}
\end{equation}
\end{linenomath}
where the sum runs over all alignment parameters and
$a_{k,\text{survey}}$ and $\sigma_{k,\text{survey}}$ represent the
survey information and its uncertainty, respectively. In this
formulation we have assumed that the survey constraints are
uncorrelated. In practise, a proper treatment of correlations is
necessary if one wants to exploit the fact that the survey uncertainty
depends on assembly granularity. For instance, the position of
detector modules in a layer or box assembly is usually much better
constrained than the position of that assembly in the global reference
frame.

Weak modes related to global distortions pose a particular concern
because they can lead to biases in track parameters that affect the
performance of an experiment. A global translation of the entire
detection apparatus changes the numerical values of the track
parameters without changing the kinematics relevant for physics
analysis, such as an invariant mass or a decay angle. However, global
distortions that affect the relative position or direction of tracks
will introduce a bias in kinematic observables and degrade the overall
detector resolution.

\begin{figure}[htb]
\includegraphics[width=.5\columnwidth]{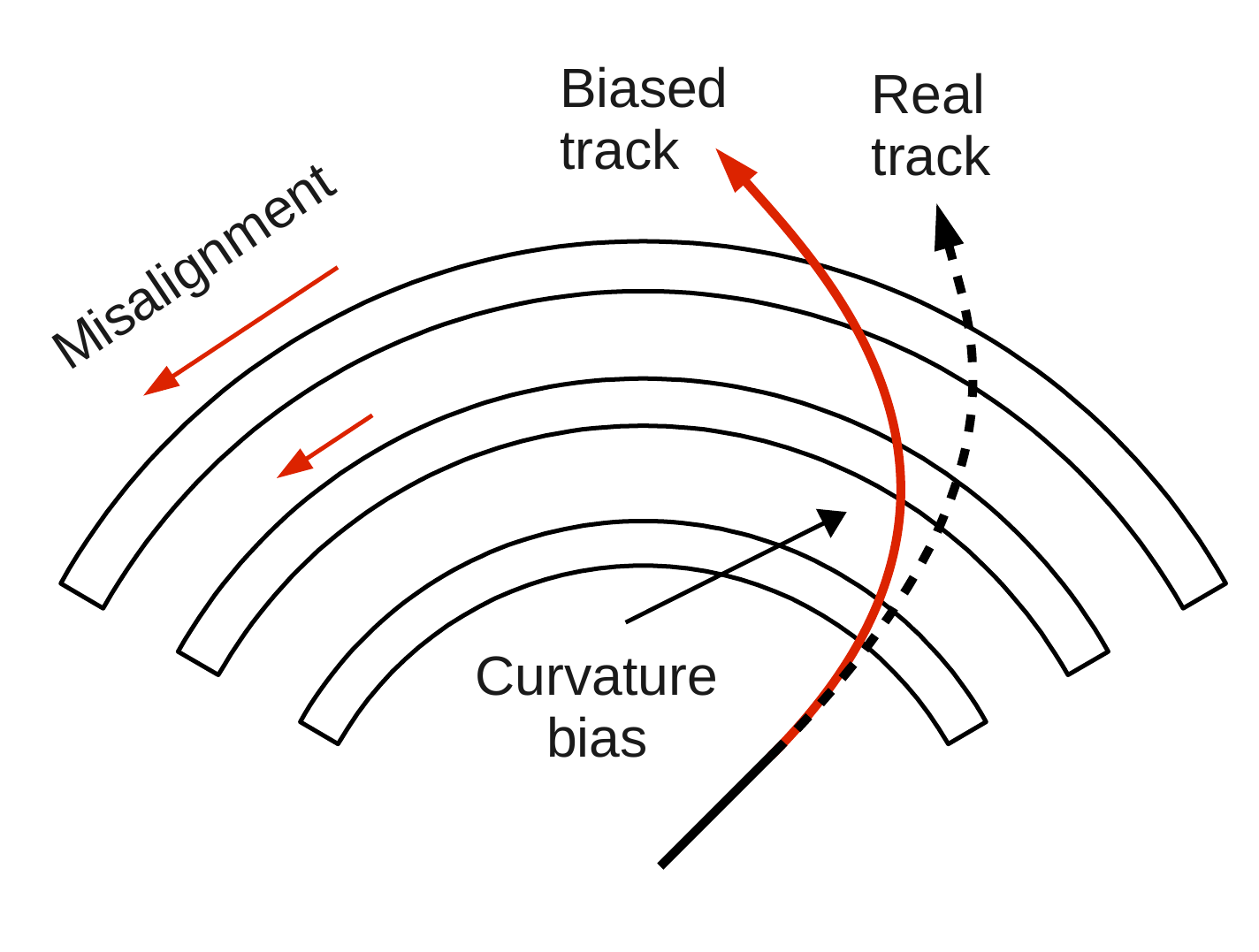}
\includegraphics[width=.5\columnwidth]{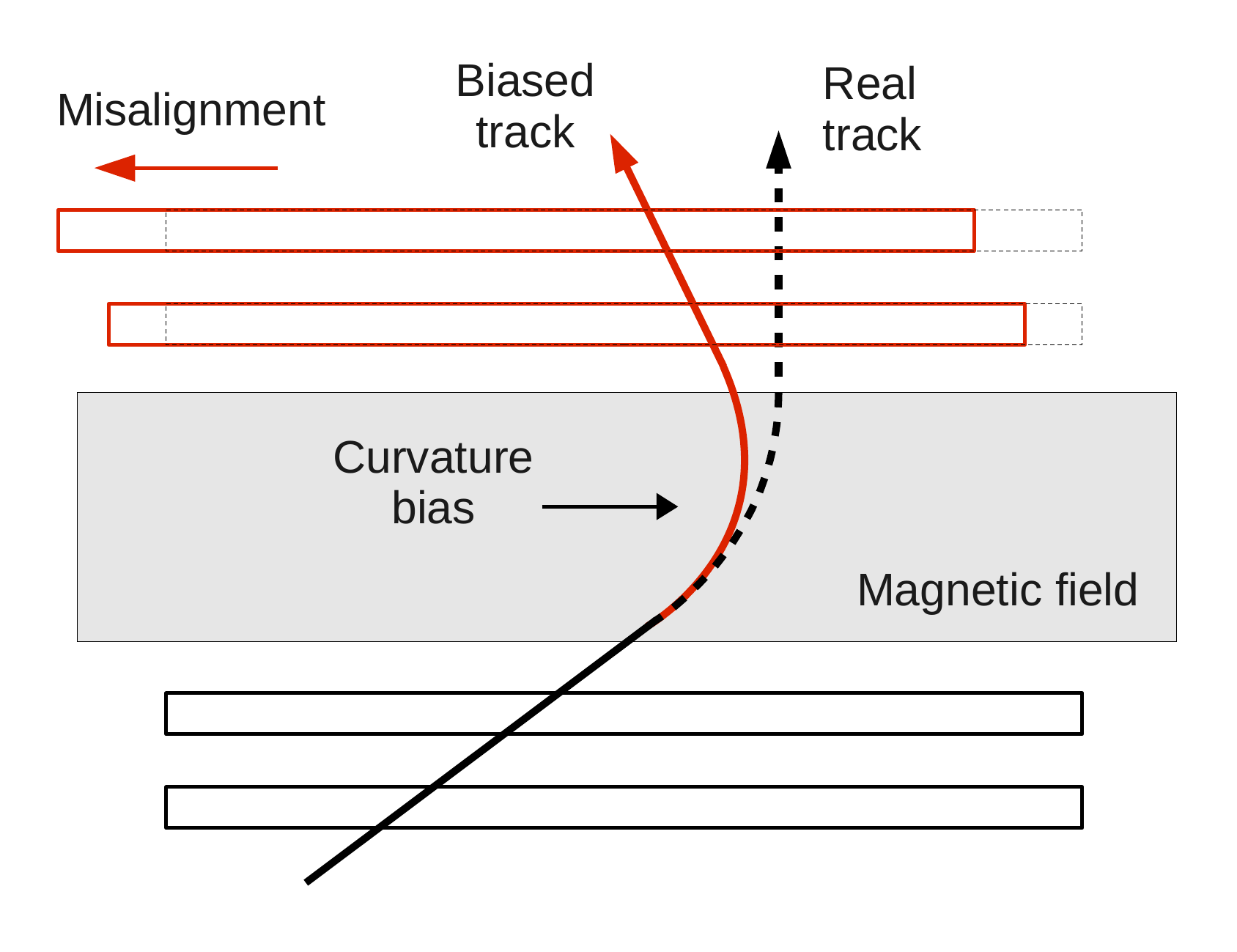}
  \caption{Illustration of a curvature bias in a cylindrical 
    detector geometry (left) and in a forward detector geometry (right).}
  \label{fig:curvaturebias}
\end{figure}

A particularly interesting weak mode related to a global distortion is
the so-called \emph{curvature bias}. This weak mode appears both in
cylindrical detectors with a solenoidal magnetic field (in which it is
sometimes called \emph{sagitta bias} or \emph{curl}) and in forward
spectrometers with a dipole magnet, as illustrated in
Fig.~\ref{fig:curvaturebias}. In cylindrical detectors it is caused by
a layer-dependent rotation. In forward detectors it can be the result
of both a relative shearing and a relative rotation of the detectors
before and after the magnet, which are to first order
indistinguishable.

The momentum of a charged particle is measured via its curvature
radius, which requires at least three measured coordinates. In a
uniform magnetic field $B$, the inverse of the curvature radius, which
we shall call the curvature $\omega$, is related to the momentum
component $p_\perp$ perpendicular to the field by
\begin{linenomath}
\begin{equation}
  \omega \; = \; \frac{Q B}{p_\perp},
\end{equation}
\end{linenomath}
where $Q$ is the charge of the particle. The weak modes illustrated in
Fig.~\ref{fig:curvaturebias} introduce a bias in the curvature
\begin{linenomath}
\begin{equation}
  \omega \; \longrightarrow \; \omega + \domega
  \label{equ:defcurvaturebias}
\end{equation}
\end{linenomath}
that for sufficiently large curvature radius is approximately
independent of particle momentum and direction. A constant curvature
bias leads to a momentum bias that depends both on momentum and
charge. As we shall see later, one signature of a curvature bias is a
shift in the reconstructed invariant mass of a two-body decay that is
proportional to the momentum difference between the two final-state
particles.

The weak mode that leads to a curvature bias only exists in the
presence of a magnetic field. Consequently, a curvature bias can be
constrained with field-off data, provided that detectors do not move
if the field is turned on. Unfortunately, the latter condition is not
often fulfilled.  In central detectors a curvature bias can be
constrained with samples of reconstructed cosmic ray tracks that
traverse detector layers on either side of their point of closest
approach to the detector axis~\cite{Brown:2008ccb}. The curvature bias
affects the `top' and `bottom' segments of such a track with opposite
signs, and hence the requirement that those curvatures must be
identical constrains the bias.  In forward detectors, such as the LHCb
detector at CERN~\cite{Alves:2008zz}, this technique does not work and
alternative methods must be deployed.

We report here on a novel method to constrain the curvature bias by
including a $\chi^2$ contribution from a mass-constrained vertex fit
of a multi-body decay. Implementations of kinematic and vertex
constraints for alignment have been presented
before~\cite{Brown:2008ccb,Flucke:1406822,Widl:2010zz,BruckmandeRenstrom:2005ha}.
With the exception of that in Ref.~\cite{BruckmandeRenstrom:2005ha},
these implementations rely on reparameterizations of the tracks and
are limited to two-track combinations. Our method does not need a
special track parametrization or track fit and can be used with
vertices with any number of tracks.

The outline of this paper is as follows. In
Section~\ref{sec:alignprocedure} we briefly discuss the track-based
alignment procedure, referring the reader to a previous publication
for details. In Section~\ref{sec:vertexfit} we present the
implementation of a vertex fit for use in the alignment. In
Section~\ref{sec:lhcbalignment} we discuss the use of vertex
constraints in the alignment of the LHCb detector. As an illustration
we compare the performance of alignments obtained with and without
mass-constrained \DzToKPi{} decays.\footnote{Throughout this paper
  charge-conjugated modes are implied.}

\section{Alignment with tracks and vertices \label{sec:alignprocedure}}

The application of a $\chi^2$ minimization procedure for track-based
alignment is extensively discussed in the literature (see
e.g.\ Ref.~\cite{Blobel:2006yh} and references therein). For our discussion
we use the formalism presented in a previous
publication~\cite{Hulsbergen:2008yv}, in which we have shown how to
use a Kalman-filter track fit in a closed-form alignment procedure. We
follow the notation from that paper and from a seminal paper on the
application of Kalman filters in high energy
physics~\cite{Fruhwirth:1987fm}.

We denote the set of alignment parameters by $a$ and the
parameters of track $i$ by $\xtrk_i$. For a track with $N$ hits, we
write the $N$ dimensional vector of hit residuals schematically as
\begin{linenomath}
\begin{equation}
  r_i( \xtrk_i, a ) \; = \; m_i( a ) - h_i( \xtrk_i ) \: ,
  \label{equ:hitresidual}
\end{equation}
\end{linenomath}
where $m_i$ is the vector of hit coordinates, each of which is a
function of (a subset of) the alignment parameters. The function $h_i$
is often called the measurement model. It expresses the expected hit
coordinates in terms of the track parameters.\footnote{Note that one
  can choose where to put the alignment parameters $a$ in
  Eq.~\ref{equ:hitresidual}. The residual can also be written as $r_i
  \; = \; m_i - h_i( \xtrk_i,a)$~\cite{Hulsbergen:2008yv}. Only the
  residual and the derivatives of the residual to track and alignment
  parameters enter the formalism.}  The $\chi^2$ of track $i$ can now
be written in matrix notation as
\begin{linenomath}
\begin{equation}
  \chi^2_i \; = \; r_i^T \, V_i^{-1} \, r_i \: ,
\end{equation}
\end{linenomath}
where $V_i$ is the \mbox{$N\!\times\! N$} covariance matrix of the
measurement coordinates. The latter is usually diagonal. The
best-fitting track parameters are obtained by minimizing $\chi_i^2$
with respect to $\xtrk_i$ for a given value $a$ of the alignment
parameters. Since the measurements depend on $a$, so do the track
parameters.

While the track parameters are different for each track, the alignment
parameters are common. To obtain the optimal alignment one minimizes
the total $\chi^2$ of a sample of tracks simultaneously with respect
to the track parameters ${x_i}$ and the alignment parameters $a$. By
taking into account how the track parameters depend on the alignment
parameters, the problem can be reduced to a minimization problem with
the dimension of $a$
only~\cite{Blobel:2002ax,Bruckman:835270,Bocci:2007zzb,Hulsbergen:2008yv}.

Starting from an initial alignment $a_0$, the solution for $a=a_0 +
\Delta a$ is obtained by solving the set of linear equations
\begin{linenomath}
\begin{equation}
  \left.\frac{\ud^2 \chi^2}{\ud a^2}\right|_{a_0} \Delta a \; = \; - \left.\frac{\ud \chi^2}{\ud a}\right|_{a_0}.
  \label{equ:deltaalignment}
\end{equation}
\end{linenomath}
The first and second derivatives are obtained by summing the
contributions from all tracks and can be expressed
as~\cite{Hulsbergen:2008yv,Bocci:2007zzb}
\newcommand{\pderiv}[2]{\frac{\partial #1}{\partial #2}}
\begin{linenomath}
\begin{equation}
  \frac{\ud \chi^2}{\ud a} \; = \; 2 \sum_\text{tracks $i$} \pderiv{r_i}{a}^{T} \: V_i^{-1} \:r_i
  \label{equ:firstalignderivative}
\end{equation}
\end{linenomath}
and
\begin{linenomath}
\begin{equation}
  \frac{\ud^2 \chi^2}{\ud a^2} \; = \; 2 \sum_\text{tracks $i$} 
      \pderiv{r_i}{a}^{T} \: V_i^{-1} \: R_i \: V_i^{-1} \: \pderiv{r_i}{a} \: ,
 \label{equ:secondalignderivative}
\end{equation}
\end{linenomath}
where $R_i$ is the covariance matrix of the residuals after the track
fit. The latter is expressed as
\begin{linenomath}
\begin{equation}
  R_i \; = \; V_i \, - \, H_i \, C_i \, H_i^T \: ,
\end{equation}
\end{linenomath}
where $H_i$ is the derivative of $r_i$ with respect to the track parameters of
track $i$ and $C_i$ is the covariance matrix for the track parameters. To
obtain the expression in Eq.~\ref{equ:firstalignderivative} one
exploits the fact that the $\chi^2$ contribution for each track has been
minimized with respect to the track parameters for the initial alignment
$a_0$~\cite{Hulsbergen:2008yv,Bocci:2007zzb}.

One ingredient to Eqs.~\ref{equ:firstalignderivative}
and~\ref{equ:secondalignderivative} is the derivatives of the
residuals (or measured hit coordinates) to alignment
parameters.\footnote{We denote these as partial derivatives since at
  this stage one ignores the contribution to the derivative that comes
  through the track parameters.} Their computation depends on the
implementation of the detector geometry and is outside the scope of
this paper.\footnote{Their evaluation in LHCb is similar to that
  discussed in Section 3 of Ref.~\cite{Bocci:2007zzb} for the ATLAS
  detector.} Another ingredient is the track parameter covariance
matrix $C_i$. In LHCb we use a Kalman filter track fit that takes
multiple scattering and energy loss into account and follow the
approach derived in Ref.~\cite{Hulsbergen:2008yv} to compute $C_i$.

In Section 4 of Ref.~\cite{Hulsbergen:2008yv} the formalism above is
extended with vertex constraints. The proposed method relies on a
vertex fit. The vertex fit computes new track parameters on the
assumption that all tracks in the fit originate at a common
point. Using the covariance matrix of the tracks the difference
between the track parameters before and after the vertex fit can be
propagated to the hit residuals $r_i$. The vertex fit introduces a
correlation between the parameters of different tracks. Therefore, the
$\chi^2$ contributions from different tracks are no longer
independent. This means that the residual vector $r_i$ now spans
residuals from all tracks included in the vertex and the covariance
matrix $C_i$ is now the covariance matrix for the parameters of all
tracks included in the vertex. The method is valid for any number of
tracks in a vertex. The strength and novel aspect of the method is
that no special track fit is required. The one ingredient that is
missing in the discussion in Ref.~\cite{Hulsbergen:2008yv} is the
vertex fit itself. We present the vertex fit in the next section.

The actual alignment procedure now consists of the following
steps. First, the tracks are fitted. Next, subsets of tracks that are
identified to come from a single vertex are combined with a vertex
fit. Using the formalism in Ref.~\cite{Hulsbergen:2008yv} a single
residual vector $r_i$ and corresponding covariance matrix $R_i$ are
computed for each multi-track object and added to the derivatives in
Eqs.~\ref{equ:firstalignderivative}
and~\ref{equ:secondalignderivative}. Selected tracks that are not used
in a vertex can also be added to these derivatives. Finally, once all
contributions in the sample have been accumulated, new alignment
parameters are computed using Eq.~\ref{equ:deltaalignment}.

\section{The vertex fit \label{sec:vertexfit}}

A vertex fit combines the trajectories of a set of charged particles
with the constraint that the particles originate from a common
point. The input to the vertex fit is the reconstructed parameters
and the covariance matrix for each of the tracks. The output of the vertex
fit is the vertex position, the momentum vector for each of the tracks and
the corresponding covariance matrix.

For the implementation of the vertex fit we use the
Billoir-Fr\"uhwirth-Regler algorithm~\cite{Billoir:1985nq}. We follow
the notation used in Ref.~\cite{Fruhwirth:1987fm}, with small
modifications to remain consistent with the symbols used for the track
parameters in the previous section. For the application in alignment
we extend the formalism with a mass constraint.

Tracks are locally parametrized by a 5-D vector, generically denoted
by the symbol $\xtrk_i$ where the label $i$ enumerates the tracks in
the vertex. The covariance matrix of the track is denoted by $C_i$ and
its inverse by $G_i$. We denote the 3-D vertex position vector with
the symbol $\xvtx$ and the 3-D vector that parametrizes the momentum
vector of outgoing track $i$ with $q_i$. The measurement model,
$h_i(\xvtx,q_i)$ expresses the parameters of track $i$ in terms of
$\xvtx$ and $q_i$. The residual of track $i$ is then defined as
\begin{linenomath}
\begin{equation}
  r_i = \xtrk_i - h_i(\xvtx,q_i) \: .
\end{equation}
\end{linenomath}

Note that the symbol $\xtrk_i$ used for the track parameters is the
same symbol that appears in Eq.~\ref{equ:hitresidual} and beyond. However,
the role of the track parameters is different: in the track fit, the
track parameters are free parameters determined by the track fit. In
the vertex fit, the track parameters and their covariance matrix are
input to the fit. In the following we show how the track parameters
are changed if a vertex constraint is applied.

The $\chi^2$ of the vertex fit is written as
\begin{linenomath}
\begin{equation}
  \chi^2 \; = \; \sum_i r_i^T \, G_i \, r_i \: ,
  \label{equ:vertexchisq}
\end{equation}
\end{linenomath}
where the sum runs over all tracks in the vertex. The solution to the
vertex fit is the set of parameters $(\xvtx, q_1,\ldots, q_N)$ that
minimizes this $\chi^2$.

As in Refs.~\cite{Fruhwirth:1987fm,Billoir:1985nq} we linearize the
measurement model around the current estimate $(\xvtx_0,q_{i,0})$,
\begin{linenomath}
\begin{equation}
  h_i( \xvtx, q_i) \; = \; h_i(\xvtx_0,q_{i,0}) \: + \: A_i ( \xvtx - \xvtx_0 ) \: + \: B_i (q_i - q_{i,0}) \: .
\end{equation}
\end{linenomath}
The measurement model and its derivatives $A_i$ and $B_i$ follow from the
parametrization chosen for tracks and vertices. In
\ref{sec:vertexmodel} we shall discuss a definition suitable for a
forward detector as LHCb. The total derivatives of the $\chi^2$ with
respect to the vertex position can now be written
as~\cite{Billoir:1985nq}
\begin{linenomath}
\begin{equation}
  \frac{\ud \chi^2}{\ud \xvtx} \; = \; - 2 \sum_i A_i^T \, \GB_i \, r_i
  \qquad\text{and}\qquad
  \frac{\ud^2 \chi^2}{\ud \xvtx^2} \; = \; 2 \sum_i A_i^T \, \GB_i \, A_i \: ,
\end{equation}
\end{linenomath}
where we introduced
\begin{linenomath}
\begin{equation}
  \GB_i \; \equiv \; G_i - G_i B_i W_i B_i^T G_i
\end{equation}
\end{linenomath}
with 
\begin{linenomath}
\begin{equation}
  W_i \; \equiv \; \left( B_i^T G_i B_i \right)^{-1}.
\end{equation}
\end{linenomath}
By requiring that the first derivative of the $\chi^2$ is zero,
updated track and vertex parameters are obtained. The vertex
parameters can be expressed as
\begin{linenomath}
\begin{equation}
  \xvtx \; = \; \xvtx_0 - \left( \frac{\ud^2 \chi^2}{\ud \xvtx^2} \right)^{-1} \frac{\ud \chi^2}{\ud \xvtx} \: ,
  \label{equ:positionupdate}
\end{equation}
\end{linenomath}
while their covariance matrix is given by
\begin{linenomath}
\begin{equation}
  \text{Cov}(\xvtx) \; \equiv \; \Cvtx \; = \; 2 \left( \frac{\ud^2 \chi^2}{\ud \xvtx^2} \right)^{-1}.
\end{equation}
\end{linenomath}
The momentum parameters of the outgoing tracks can be computed with
\begin{linenomath}
\begin{equation}
  q_i \; = \; q_{i,0} + W_i B_i^T G_i ( \xtrk_i - h_i(\xvtx, q_{i,0}) ) \: .
  \label{equ:momentumupdate}
\end{equation}
\end{linenomath}
The covariance matrix for two outgoing tracks $i$ and $j$ is given by
\begin{linenomath}
\begin{equation}
  \text{Cov}(q_i,q_j) \equiv D_{i,j} \; = \; \delta_{ij} W_i \: + \: (W_i B_i^T G_i A_i) \, \Cvtx \, (A_j^T G_j B_j W_j) \: ,
  \label{equ:mommomcov}
\end{equation}
\end{linenomath}
where $\delta_{ij}$ is one if $i=j$ and zero otherwise.  The
covariance matrix for the vertex position and the momentum of track
$i$ is given by
\begin{linenomath}
\begin{equation}
  \text{Cov}(q_i,\xvtx) \equiv E_i \; = \; - W_i B_i^T G_i A_i \Cvtx \: .
  \label{equ:posmomcov}
\end{equation}
\end{linenomath}

The vertex fit can be iterated until a certain convergence criterion
is met, for example a sufficiently small change in the vertex
$\chi^2$ of Eq.~\ref{equ:vertexchisq}. Note that the computation of the
covariance matrices in Eqs.~\ref{equ:mommomcov}
and~\ref{equ:posmomcov} is CPU intensive, but not needed for obtaining
the $\chi^2$ or the position and momentum updates in
Eqs.~\ref{equ:positionupdate}
and~\ref{equ:momentumupdate}~\cite{Billoir:1985nq}. Consequently,
their computation can be delayed until the fit has converged.

\newcommand{\pmu}{\ensuremath{k}}

An invariant mass constraint adds a new term to the total vertex
$\chi^2$ in Eq.~\ref{equ:vertexchisq}.  In contrast to the $\chi^2$
contributions from the individual tracks, the mass $\chi^2$ term has
non-zero derivatives to all momentum vectors $q_i$, instead of just
one of them. As a result the formulation of the vertex fit
above cannot easily be extended with a mass constraint. Therefore, we
have chosen to add the mass constraint \emph{after} the fit to the
vertex position, ignoring small effects due to the non-linearity of
the fit. Such effects can be expected to be small as long as the
change in the invariant mass is small compared to the invariant mass.

For each final state particle $i$, we compute a relativistic four-momentum
vector $\pmu_i$, which is a function of the momentum parameters $q_i$
and a mass hypothesis $m_i$. We denote the $(4\times 3)$ matrix for
the derivative of $\pmu_i$ with respect to $q_i$ by
\begin{linenomath}
\begin{equation}
  K_i \; \equiv \; \frac{\ud \pmu_i}{\ud q_i} \: .
  \label{equ:matrixK}
\end{equation}
\end{linenomath}
The total four-momentum of all tracks assigned to the vertex and its
covariance matrix are given by
\begin{linenomath}
\begin{equation}
  \pmu_\text{tot} \; = \; \sum_i \pmu_i(q_i, m_i) \qquad \text{and} \qquad
  \text{Cov}(\pmu_\text{tot}) \; = \; \sum_{i,j} \: K_i \: D_{i,j} \: K_j^T ,
\end{equation}
\end{linenomath}
where $q_i$ and $D_{i,j}$ are the result of the unconstrained fit
(Eqs.~\ref{equ:momentumupdate} and~\ref{equ:mommomcov},
respectively). From $\pmu_\text{tot}$ we compute the mass and form a
residual for the mass constraint,
\begin{linenomath}
\begin{equation}
  r_\mathrm{M} \; = \; m(\pmu_\text{tot}) - m_0 \: ,
\end{equation}
\end{linenomath}
where $m_0$ is the known mass of the reconstructed decay. Defining the
$1 \times 4$ derivative matrix
\begin{linenomath}
\begin{equation}
  H_\mathrm{M} \; \equiv \; \frac{\ud m}{\ud \pmu_\text{tot}} \: ,
  \label{equ:matrixHM}
\end{equation}
\end{linenomath}
the variance of the constraint is given by
\begin{linenomath}
\begin{equation}
  R_\mathrm{M} \; = \; H_\mathrm{M} \: \text{Cov}(\pmu_\text{tot}) \: H_\mathrm{M}^T \: .
\end{equation}
\end{linenomath}
In case the natural width $\Gamma_0$ of the decaying particle is not
small compared to the invariant mass resolution, such as for the decay
\mbox{$Z^0 \to \mumu$}, one can add $\Gamma_0^2$ to the variance of
the constraint.

Using well-known expressions for the Kalman
filter~\cite{Fruhwirth:1987fm}, the mass-constrained vertex position
and track parameters now become
\begin{linenomath}
\begin{align}
  \hat{\xvtx} & = \; \xvtx - \sum_k E_k^{T} K_k^{T} H_\mathrm{M}^T R_\mathrm{M}^{-1} r_\mathrm{M} \: ,
  \label{equ:massconstraintpars}\\
  \hat{q}_i  & = \; q_i -  \sum_{k} D_{i,k} K_k^{T} H_\mathrm{M}^T R_\mathrm{M}^{-1} r_\mathrm{M} \: .
\end{align}
\end{linenomath}
The updated covariance matrices are given by
\begin{linenomath}
\begin{align}
  \widehat{C}_\xvtx & = \; \Cvtx - \sum_{k,l} E_k^{T} K_k^{T} H_\mathrm{M}^T R_\mathrm{M}^{-1} H_\mathrm{M} K_l E_l \:,\\
  \widehat{D}_{i,j} & = \; D_{i,j} - \sum_{k,l} D_{i,k} K_k^{T} H_\mathrm{M}^T R_\mathrm{M}^{-1} H_\mathrm{M} K_l D_{l,j} \:, \\
  \widehat{E}_i    & = \; E_{i} - \sum_{k,l} D_{i,k} K_k^{T} H_\mathrm{M}^T R_\mathrm{M}^{-1} H_\mathrm{M} K_l E_{l} \:.
  \label{equ:massconstraintcovs}
\end{align}
\end{linenomath}
The indices $k$ and $l$ run over all tracks in the vertex.

Finally, for application in the alignment, the vertex-constrained
track parameters and their covariance matrices must be computed. These
follow from the measurement model as
\begin{linenomath}
\begin{equation}
  \tilde{\xtrk}_i \; = \; h( \xvtx, q_i)
\end{equation}
\end{linenomath}
and
\begin{linenomath}
\begin{equation}
  \text{Cov}( \tilde{\xtrk}_i, \tilde{\xtrk}_j ) \; = \; 
  A_i \Cvtx A_j^T + A_i E_j^T B_j^T + B_i E_i A_j^T  
  + B_i D_{i,j} B_j^T \: .
\end{equation}
\end{linenomath}
These equations provide the input to Eqs. 18, 19 and 21 in
Ref.~\cite{Hulsbergen:2008yv}.\footnote{The symbols
  $\tilde{x}_0^{(i)}$, $\tilde{C}_0^{(i)}$ and $\tilde{C}_0^{(i,j)}$
  in Ref.~\cite{Hulsbergen:2008yv} translate in our notation as
  $\tilde{\xtrk}_i$, $\text{Cov}( \tilde{\xtrk}_i, \tilde{\xtrk}_i)$
  and $\text{Cov}( \tilde{\xtrk}_i, \tilde{\xtrk}_j )$, respectively.}
In case a mass constraint is used we replace the vertex position
$\xvtx$, the track momentum parameters $q_i$ and their corresponding
covariance matrices with their mass-constrained counterparts in
Eqs.~\ref{equ:massconstraintpars}--\ref{equ:massconstraintcovs}. This
concludes the algebra of the vertex fit for use in a track-based
alignment algorithm.

\section{Application to the alignment of the LHCb spectrometer}
\label{sec:lhcbalignment}

\begin{figure}
  \centerline{\includegraphics[width=0.65\textwidth]{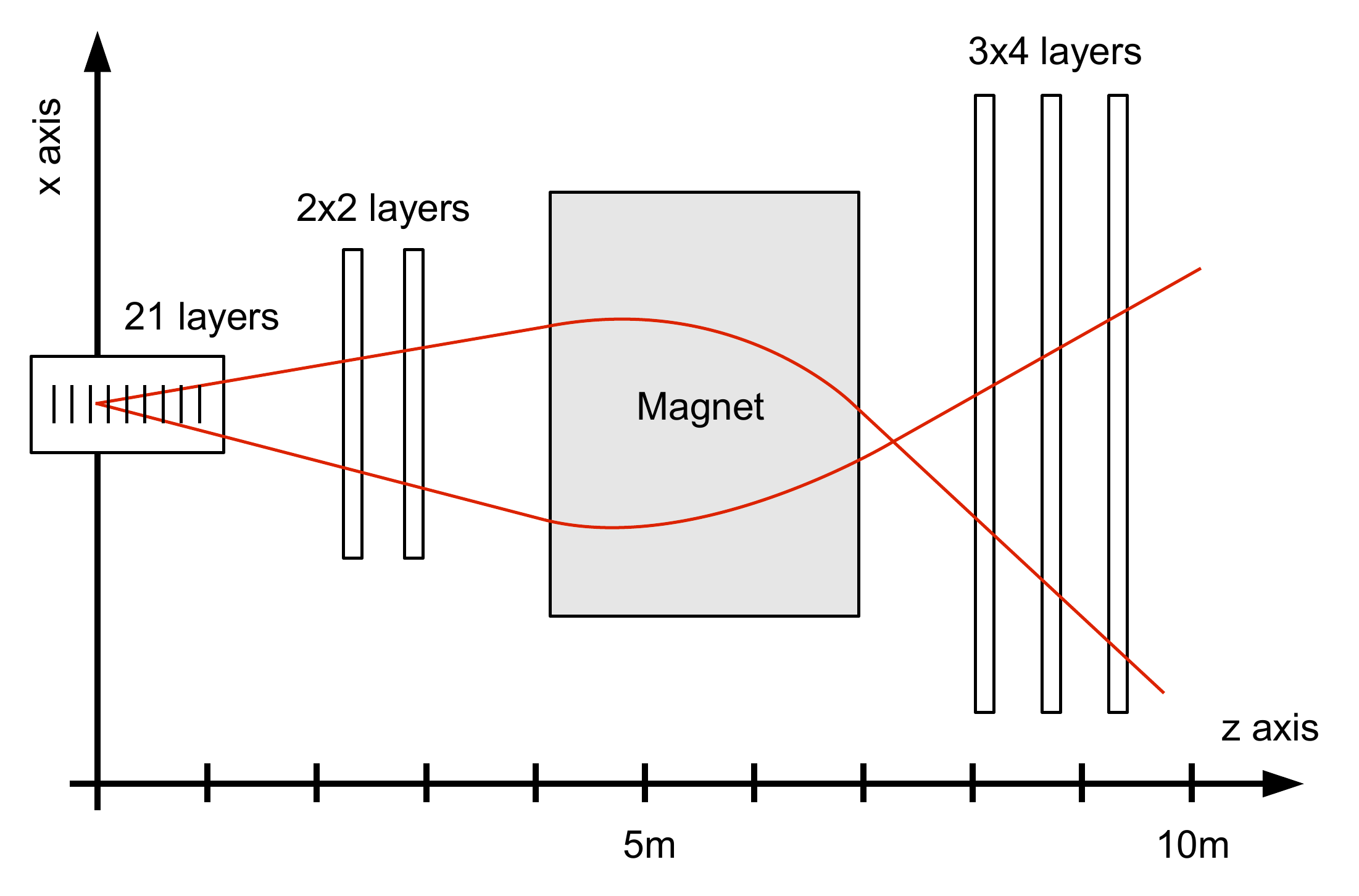}}
  \caption{Sketch of the LHCb spectrometer. The proton-proton
    collision point is located inside the vertex detector on the left
    ($z \simeq 0$). The main component of the field of the dipole
    magnet is parallel to the $y$ axis. The curved lines represent
    trajectories of a positively and negatively charged particle.}
  \label{fig:lhcbdetector}
\end{figure}

The tracking system of the LHCb detector is an example of a planar
detector with a forward geometry. It is schematically depicted in
Fig.~\ref{fig:lhcbdetector} and discussed in detail in
Ref.~\cite{Alves:2008zz}. Charged particles produced at the
interaction point bend in the magnetic field of a dipole magnet with a
field integral of about $4$~Tm. Precision vertexing is provided by a
21-layer silicon strip detector located in the field-free region
around the interaction point. Four more layers of silicon strip
detectors just in front of the magnet and another 12 layers of silicon
strip detectors and straw tube chambers behind the magnet allow for a
precise momentum measurement.

The coordinate frame in LHCb is defined such that the $y$ axis is
pointing upwards (parallel to the main component of the dipole field)
and the $z$ axis is parallel to the beam line with positive $z$ in the
direction of the spectrometer. The $x$ axis is chosen such that
$(x,y,z)$ is a right-handed system. The origin is located
approximately in the center of the vertex detector and roughly
corresponds to the average interaction point.

The implementation of a global minimum $\chi^2$ algorithm for the
alignment of tracking detectors in LHCb has been previously discussed
in Refs.~\cite{Nicolas:2008zz,Amoraal:2010zz}. Since then we have extended
the algorithm to exploit vertex constraints from primary vertices and
from resonances using the techniques outlined above.

\subsection{Primary vertex constraints}

Primary vertices are important for the alignment of the LHCb vertex
detector to guarantee an optimal impact parameter and decay time
resolution~\cite{Viret:2008jq}. The silicon modules of the vertex
detector are assembled in two detector `halves' that are positioned on
the positive and negative $x$ side of the LHC beam line.  The fraction
of tracks leaving hits in both halves is small. Furthermore, tracks
that cross detector planes both in front and behind the average
interaction point (at $z=0$) are rare as well. Reconstructed primary
vertices allow to link detector planes at positive and negative $x$
and at positive and negative $z$.

We have successfully exploited primary vertices in LHCb alignment
using the algorithm described above. Although the algorithm in
principle allows to use vertices with an arbitrary number of tracks, a
practical problem occurs for high track multiplicity.  In LHCb primary
vertices often contain tens of reconstructed tracks. Reconstructed
tracks can have up to 40 hits. A first implementation of the algorithm
showed that for large-multiplicity primary vertices
the computation of the correlations between all hits on all tracks is
computationally very demanding. In Ref.~\cite{Hulsbergen:2008yv} it
was suggested to compute only the correlations between the hits
nearest to the vertex. However, we have found that this can lead to
non-positive definite contributions to the second derivative of the
$\chi^2$ and therefore is not a viable solution.

A working solution has been obtained by limiting the track
multiplicity in vertices. For application in alignment we divide a
reconstructed primary vertex into separate vertices with at most eight
tracks each. Tracks are sorted such that particles flying forward
(\mbox{$p_z >0$}) or backward (\mbox{$p_z<0$}) and left (\mbox{$p_x>0$}) or right (\mbox{$p_x<0$})
are distributed evenly over the different vertices. In our framework
the computation time of the correlations in these vertices is small
compared to the overall reconstruction and track fitting time. The
loss in statistical power due to the splitting of the vertex can be
compensated by using more events. With typically 30 tracks per primary
vertex, we need about 20\% more data to compensate for the loss.

\subsection{Invariant mass constraints}

To constrain weak degrees of freedom in the spectrometer, such as the
curvature bias discussed in the introduction, mass-constrained
vertices from \DzToKPi{}, \JpsiToMuMu{} and \mbox{$Z^0\to\mumu$}
decays are used. The advantage of \DzToKPi{} over other resonances is
their large abundance and clean secondary vertex signature in
LHCb. This allows to select samples of thousands of events per hour
with practically no background.

In order to illustrate the effect on the alignment and on momentum
measurements, we show a comparison of two alignment strategies, one
obtained using approximately $300$k selected high-momentum tracks, and
another using in addition the constraint from $80$k mass-constrained
\DzToKPi{} vertices in the same sample.  For this exercise all
alignment parameters for the LHCb vertex detector were fixed, while
all detector elements behind the vertex detector were allowed to move
in the $x$ direction and rotate in the $xy$ plane around their center
of gravity.  To constrain weak modes and ensure convergence, survey
information was used by adding for each alignment parameter a term to
the total $\chi^2$, as in Eq.~\ref{equ:surveychisq}. The alignment
process started in both cases from alignment parameters obtained with
early data~\cite{Nicolas:2008zz,Gersabeck:2008jr,*Borghi:2010zza}.  To
account for non-linearities multiple iterations were performed. In
each iteration the same data set was used, but the assignment of hits
to tracks, the track fit and the track selection were redone. A single
iteration took approximately 1 hour on a 2.8 GHz CPU.

\begin{figure}[htb]
\begin{center}
  \includegraphics[width=.6\columnwidth]{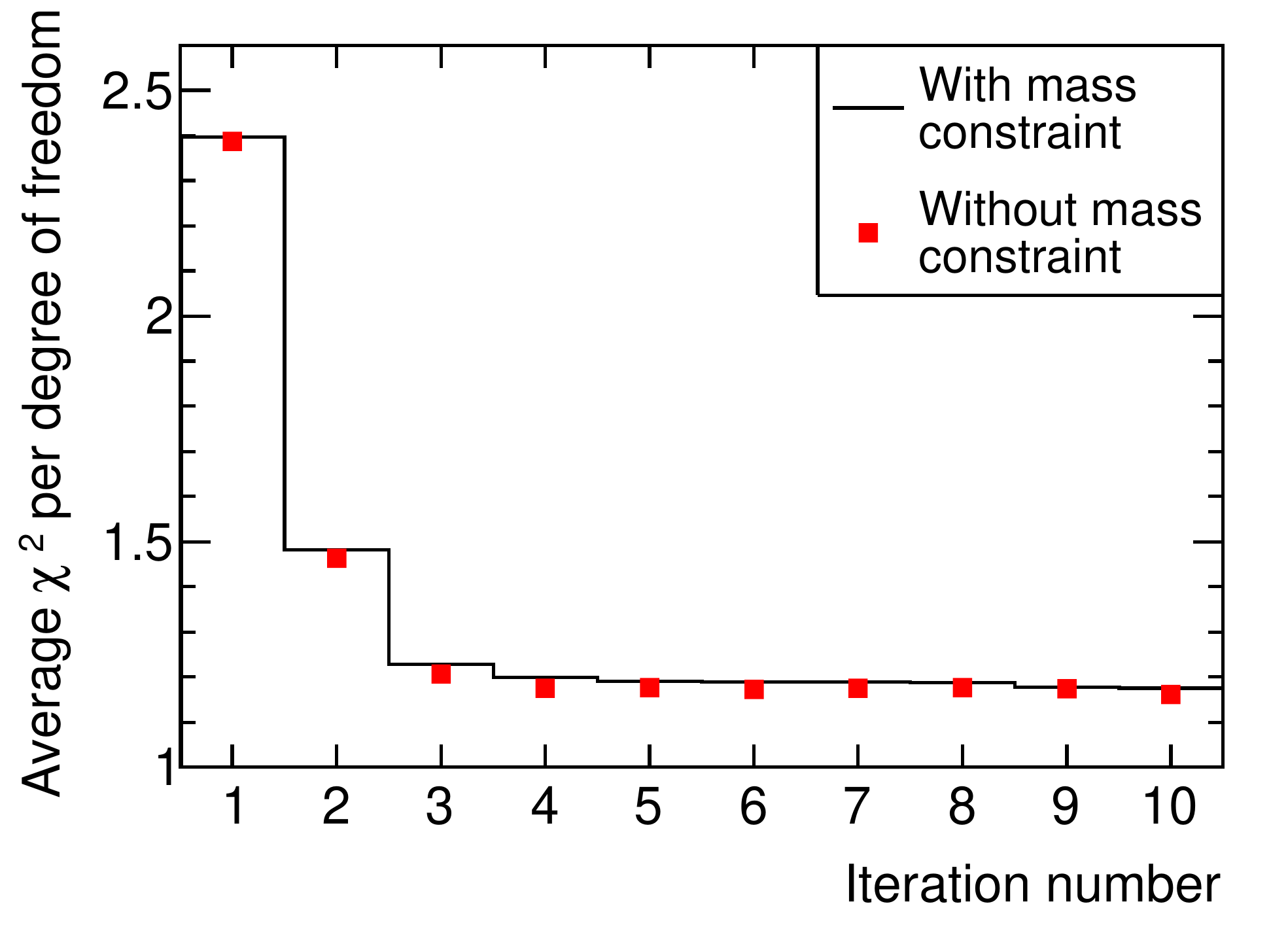}
  \caption{Average $\chi^2$ per degree of freedom at each iteration of
    the alignment procedure for the alignment with (black line) and
    without (red squares) \Dz{} mass constraint.}
  \label{fig:convergence}
\end{center}
\end{figure}

In both scenarios the minimization converged in about three iterations, as
illustrated by the average $\chi ^2$ per degree of freedom versus
iteration, shown in Fig.~\ref{fig:convergence}. Remaining variations
in the $\chi^2$ between iterations are due to small changes in the
track sample entering the alignment, as individual tracks are added or
removed. We verified the convergence by running many more iterations
and by studying the stability of alignment parameters and curvature
bias. While slowly converging components exist, these do not affect
the results reported below.

\begin{figure}[hbt]
  \includegraphics[width=.5\columnwidth]{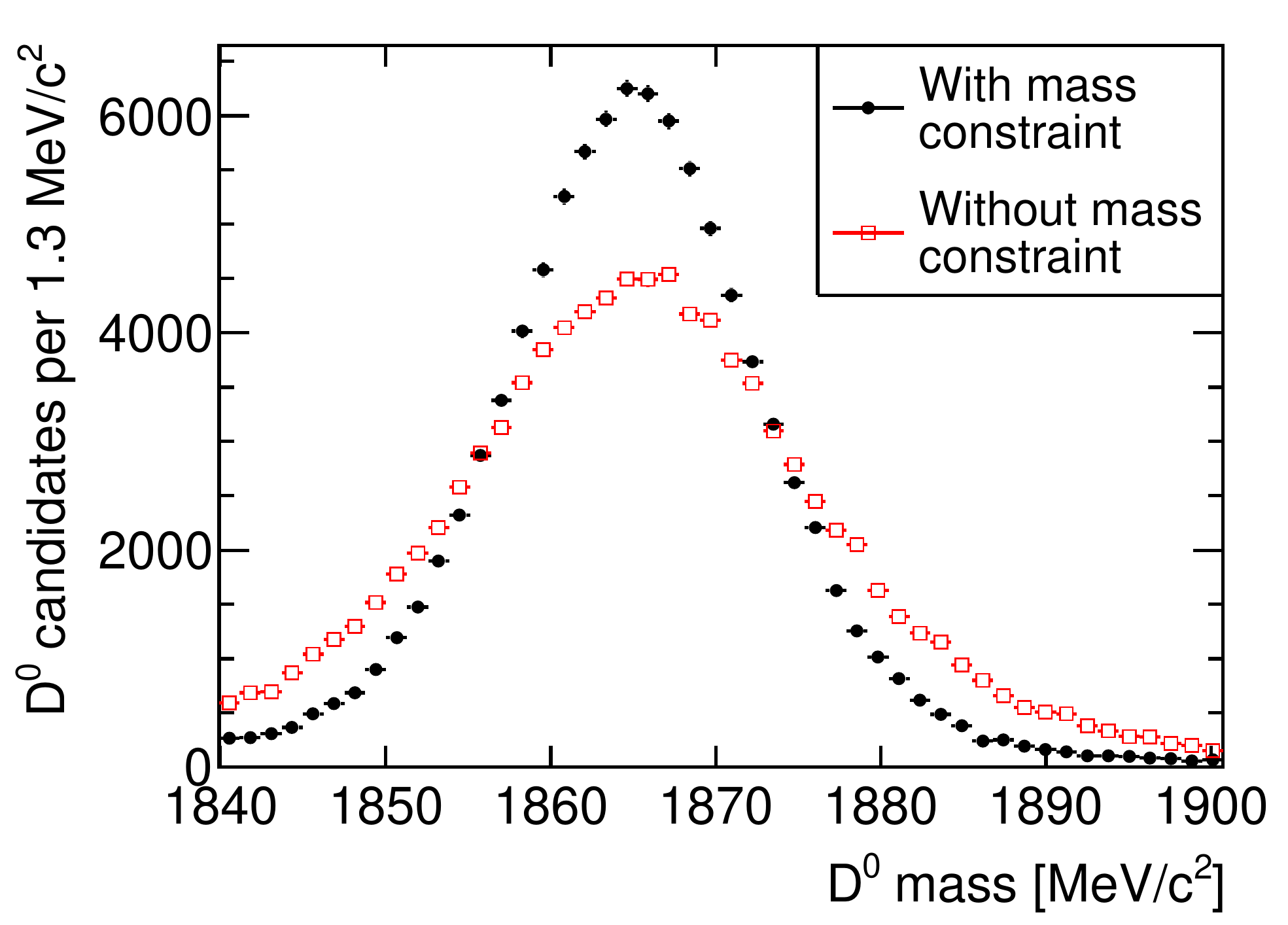}
  \includegraphics[width=.5\columnwidth]{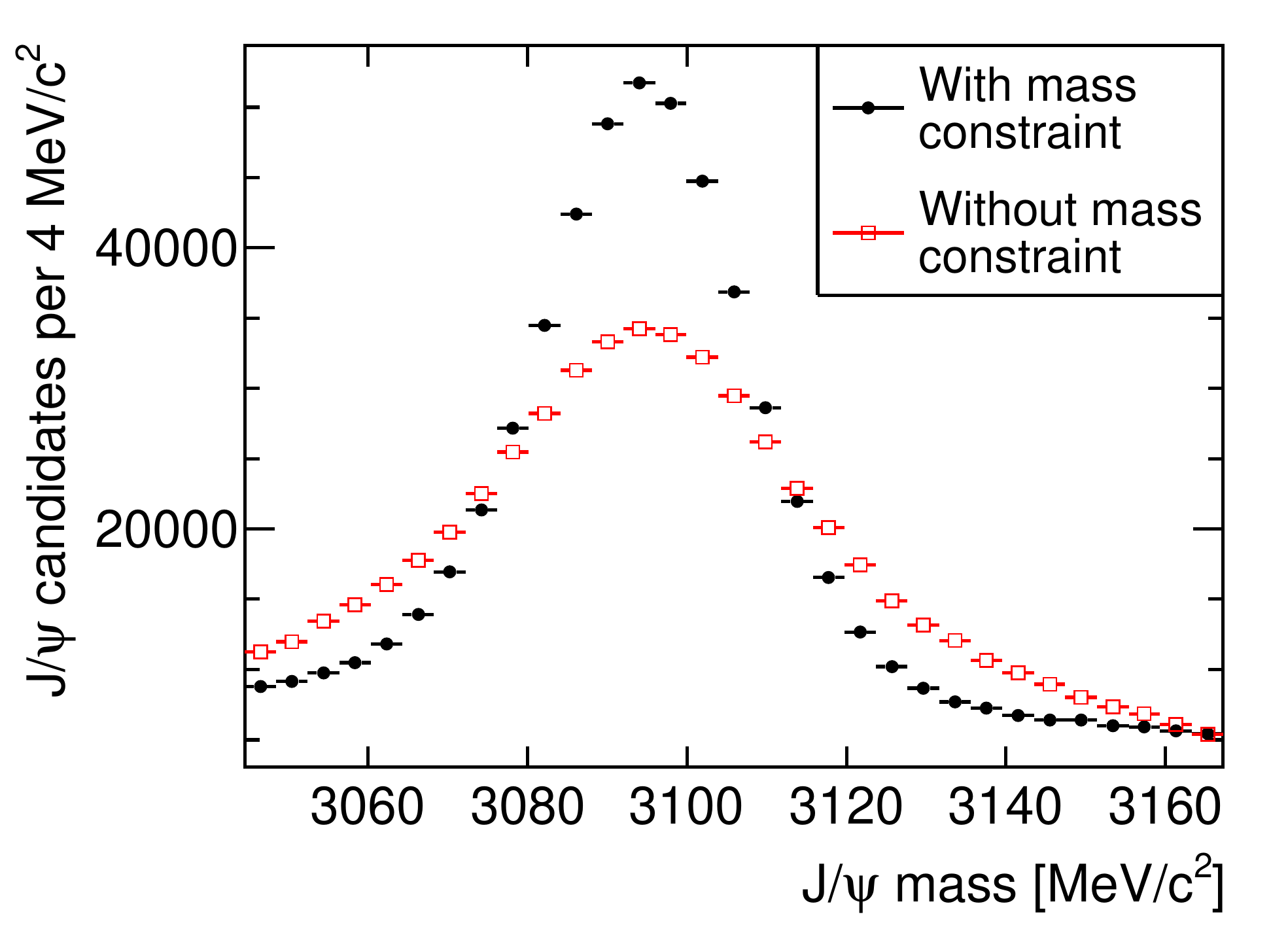}
  \caption{Invariant mass distribution of \DzToKPi{} (left) and
    \JpsiToMuMu{} (right) candidates obtained with the alignment using
    the \Dz{} mass constraint (black solid points) and the alignment
    based only on tracks (red open points).}
  \label{fig:masses}
\end{figure}

Although the average track $\chi^2$ after the alignment is practically
identical in the two cases, the performance in terms of invariant mass
resolution is very different. Figure~\ref{fig:masses} shows the
invariant mass distribution of \DzToKPi{} and \JpsiToMuMu{} candidates
on independent data sets using the two alignment sets as input. The
mass resolution obtained with the alignment that exploits the
\DzToKPi{} mass constraint is approximately 30\% better.

To understand this behaviour we consider the effect of a curvature
bias on the invariant mass of a decay to two oppositely charged
particles with momenta $p_-$ and $p_+$ and masses $m_-$ and $m_+$,
respectively. Assuming small masses with respect to the momenta,
i.e.\ ignoring terms of order $m_i^2 /p_i^2$, the invariant mass is
given by
\begin{linenomath}
\begin{equation}
  m \; = \; \sqrt{ m_-^2 + m_+^2 + 2 p_- p_+ (1 - \cos\theta) / c^2 } \: ,
\end{equation}
\end{linenomath}
where $\theta$ is the opening angle between the two particles. As a
result of a curvature bias $\domega$, the momentum changes as
\begin{linenomath}
\begin{equation}
  p_\perp \longrightarrow 
  p_\perp \left( 1 - \frac{\domega}{\omega} \right) \; = \; 
  p_\perp \left( 1 - \frac{\domega p_\perp}{Q B} \right) \: ,
\end{equation}
\end{linenomath}
where we have ignored higher order terms in $\domega/\omega$. Note
that the sign of the bias in the momentum is opposite for the positive
and the negative track. In a forward spectrometer the momentum
component perpendicular to the field dominates the total momentum. The
change in the invariant mass then becomes, to first order in
$\domega/\omega$,
\begin{linenomath}
\begin{equation}
\label{equ:massbias}
m \; \longrightarrow \; m \: \left ( 1 \: - \: (p_+ - p_-) \: \frac{m^2 - m_-^2 - m_+^2}{2m^2} \: \frac{ \domega }{e B} \right) \: ,
\end{equation}
\end{linenomath}
where $e$ is the positron charge. In other words, one expects a bias
in the mass that is approximately proportional to the difference of
the momenta of the two final state particles.

\begin{figure}[htb]
  \includegraphics[width=.5\columnwidth]{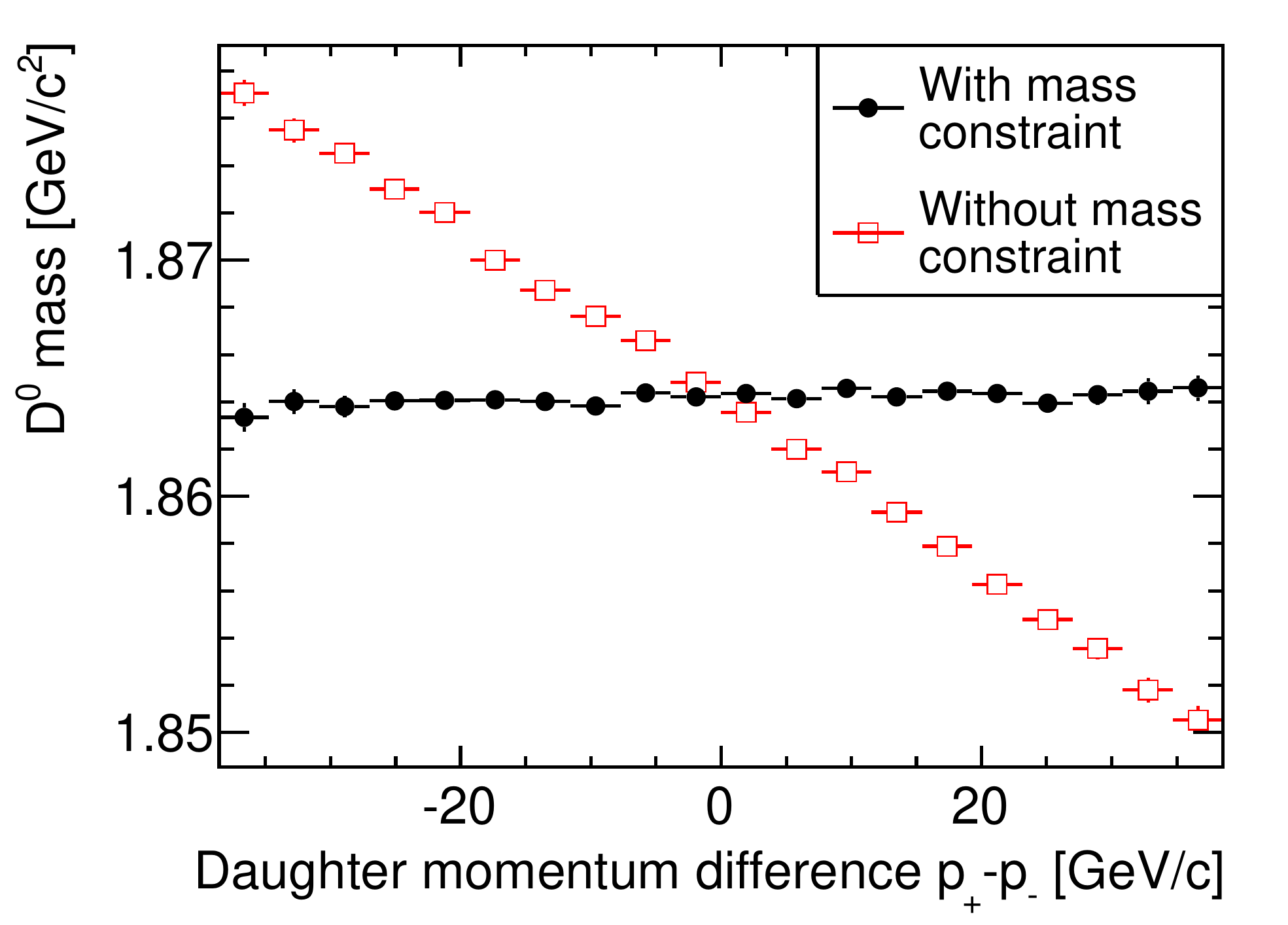}
  \includegraphics[width=.5\columnwidth]{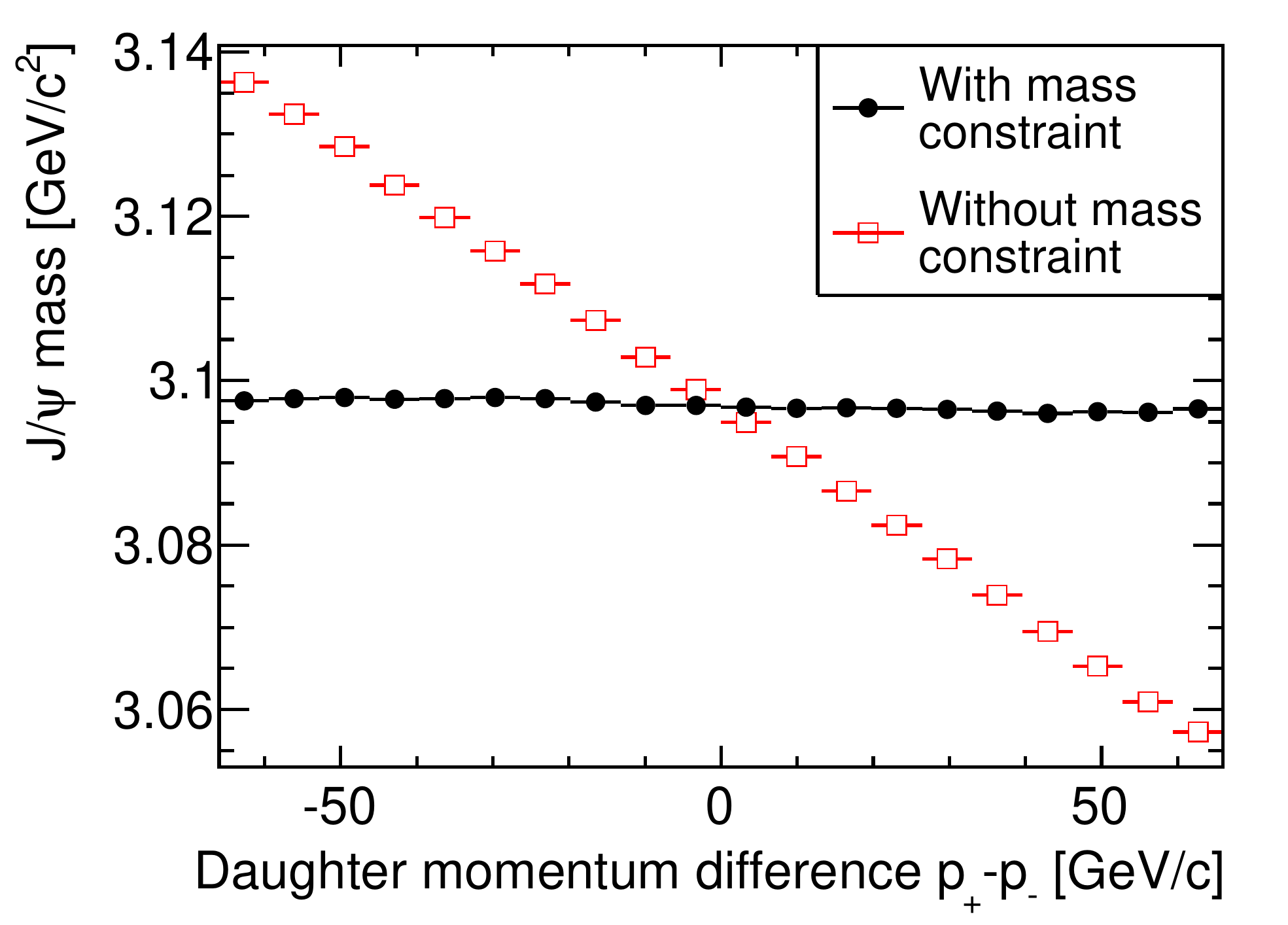}
  \caption{Position of the peak of the invariant mass distribution for
    \DzToKPi{} (left) and \JpsiToMuMu{} (right) candidates as a
    function of the momentum difference of the two daughter tracks
    evaluated with the alignment using the \Dz{} mass constraint and
    the alignment based only on tracks.}
  \label{fig:momdif}
\end{figure}

This effect is demonstrated in Fig.~\ref{fig:momdif}. The figure shows
the position of the peak of the invariant mass distribution for
\DzToKPi{} and \JpsiToMuMu{} decays as a function of the momentum
difference of the final state particles in the two scenarios. Indeed,
if the mass constraint is not used a clear evidence of a curvature
bias is observed. The removal of the curvature bias explains the
difference in mass resolutions shown in Fig.~\ref{fig:masses}.

From the slopes of the graphs in Fig.~\ref{fig:momdif} one can
estimate the value of $\domega / 2 e B$. We verify that the two slopes
are consistent with a single curvature bias. Studies with simulated
LHCb data documented in Ref.~\cite{Simioni:thesis} show that the curvature
bias is proportional to the $x$ displacement of the detector planes
behind the magnet. The reported constant of proportionality allows us
to predict the displacement from the observed slopes in
Fig.~\ref{fig:momdif}. We compute a value of approximately 1.8~mm for
the $x$ movement of the tracking layers behind the magnet, consistent
with the actually observed difference in alignment constants between
the scenarios with and without mass constraint.

As explained in the introduction, a curvature bias may appear in the
scenario without the mass constraint because it corresponds to a weak
mode, a common displacement of tracking layers that is not constrained
by single particle trajectories. Weak modes are constrained in the
LHCb alignment by survey information. The precision of the survey is
poor compared to the resolution of individual tracks. However, since
tracks carry little information on the weak modes, in the absence of mass
constraints the precision of the survey limits the precision to which
the curvature bias can be constrained.  The 1.8~mm difference between
the two alignments seems large compared to the typical uncertainty of
the survey information for the detector layers behind the magnet,
which is about 0.5~mm. However, for the results obtained here the
vertex detector was used as a reference. The observed displacement is
equivalent to a 0.2~mrad rotation of the vertex detector around its
$y$ axis, consistent with the precision of the vertex detector survey.

\section{Conclusion}

Following a recipe outlined in Ref.~\cite{Hulsbergen:2008yv} we have
developed a method to extend a track-based minimum $\chi^2$ algorithm
for detector alignment with information from reconstructed
vertices. We have presented the algebra of the vertex fit, including a
mass constraint. We have demonstrated how such a vertex fit can be
exploited to remove the effect of a curvature bias in the alignment
procedure.

Primary vertex constraints and $\DzToKPi$ mass constraints are now by
default applied in the LHCb alignment procedure.  For LHCb, invariant
mass resolution is important to isolate rare $B$ decays from the
background and to separate decays that are kinematically close, such
as decays of $B$ mesons to two light hadrons. The excellent mass
resolution of the spectrometer allows the LHCb collaboration to
perform world-best measurements in such
decays~\cite{Aaij:2012qe,*Aaij:2012ac}.

\section*{Acknowledgements}

\noindent The authors of this paper are members and ex-members of the
LHCb collaboration. For the analysis presented in
Section~\ref{sec:lhcbalignment} we are indebted to the LHCb collaboration
for the use of the LHCb software environment and LHCb data. Individual
authors acknowledge support from various National Agencies, in
particular CERN, CNRS/IN2P3 (France), BMBF, DFG, HGF and MPG
(Germany), INFN (Italy), FOM and NWO (The Netherlands), SNSF and SER
(Switzerland), STFC (United Kingdom) and NSF (USA).

\appendix

\section{Track and vertex model for a forward detector}
\label{sec:vertexmodel}

Here we discuss a parametrization for the vertex fit in a forward
detector like LHCb. We define a cartesian coordinate frame as in
Section~\ref{sec:lhcbalignment}. Track trajectories are locally
parametrized by a 5-D `state' vector $(x,y,t_x,t_y,\kappa)$ at a
given $z$ coordinate, such that $(x,y,z)$ is a point on the track,
$t_x$ and $t_y$ are the local tangents $\ud x/\ud z$ and $\ud y/\ud
z$, respectively, and $\kappa = Q/(pc)$ with $Q$ the charge and $p$ the
momentum.

In the vertex fit the vertex coordinates $\xi$ are parametrized by a
vector $\xi = (x_v, y_v, z_v)$. The momentum vector of each of the
outgoing tracks $i$ is parametrized by a 3-D vector $q_i = (t_{x,v,i},
t_{y,v,i}, \kappa_{v,i})$, where the meaning of parameters is the same
as in the measured track state. The subscript $v$ indicates that these
are now parameters of the fit, not reconstructed track parameters. The
motivation for choosing this particular parametrization is that it
makes the measurement model nearly linear.

We assume that the reference $z$ coordinate of each of the track
states is sufficiently close to the $z$ coordinate of the vertex that
the magnetic field can be ignored. The measurement model for track $i$
then becomes
\begin{linenomath}
\begin{equation}
  h( \xi, q_i ) \; = \; \left(
    \begin{array}{c}
      x_v + (z_i - z_v) t_{x,v,i} \\
      y_v + (z_i - z_v) t_{y,v,i} \\
      t_{x,v,i} \\
      t_{y,v,i} \\
      \kappa_{v,i} \\
    \end{array}
    \right) \: ,
\end{equation}
\end{linenomath}
where $z_i$ is the position at which the measured state vector of the
track is defined. For the corresponding derivative matrices we obtain
\begin{linenomath}
\begin{equation}
  A( \xi, q_i ) \; = \; \left(
    \begin{array}{ccc}
      1 & 0 & - t_{x,v,i} \\
      0 & 1 & - t_{y,v,i} \\
      0 & 0 & 0 \\
      0 & 0 & 0 \\
      0 & 0 & 0 \\
    \end{array}
    \right)
\end{equation}
\end{linenomath}
and 
\begin{linenomath}
\begin{equation}
  B( \xi, q_i ) \; = \; \left(
    \begin{array}{ccc}
      z_i - z_v &  0 & 0 \\
      0 & z_i - z_v & 0 \\
      1 & 0 & 0 \\
      0 & 1 & 0 \\
      0 & 0 & 1 \\
    \end{array}
  \right).
\end{equation}
\end{linenomath}
Note that in the absence of a magnetic field, the magnitude of the
momentum is not required to obtain the vertex position. Therefore, the
fit can be performed more efficiently by omitting the curvature
parameter altogether and working with a 4-D track model. The updated
curvature can then be computed afterward by propagating the change in
the remaining four track parameters. This approach requires more
algebra, in particular when implementing the mass constraint.
Consequently, we have not used it.

For completeness we also present the expressions for the derivative
matrices needed for the mass constraint. We drop the subscript $v$ for
readability and choose units such that $c=1$. In terms of the fit
parameters the four-vector $(\vec{k}_i,k_{0,i})$ of outgoing track $i$
becomes
\begin{linenomath}
\begin{equation}
  \pmu_i \; = \; \left( 
  \begin{array}{c}
  p_i \, t_{x,i} \, / \, n_i\\
  p_i \, t_{y,i} \, / \, n_i\\
  p_i \, / n_i\\
  \sqrt{m_i^2 + p_i^2} 
  \end{array}
  \right) \: ,
\end{equation}
\end{linenomath}
where $m_i$ is the track candidate mass, $p_i = Q_i/\kappa_i$ and $n_i = \sqrt{1
  + {t_{x,i}}^2 + {t_{y,i}}^2}$. The derivative matrix in
Eq.~\ref{equ:matrixK} is then given by
\begin{linenomath}
\begin{equation}
  K_i \; = \; \left(
  \begin{array}{ccc}
    (1+ t_{y,i}^2) \, p_i \, / \, n_i^3  & - t_{x,i} \, t_{y,i} \, p_i \, / \, n_i^3 & - Q_i \, t_{x,i} \, p_i^2 \, / \, n_i\\
    - t_{x,i} \,t_{y,i} \, p_i \,/ \,n_i^3     & (1 + t_{x,i}^2) \, p_i \,/\, n_i^3 & - Q_i \,t_{y,i} \,p_i^2 \,/\, n_i \\
    -t_{x,i} \,p_i \,/ \,n_i^3            & -t_{y,i} \, p_i \, / \, n_i^3 & -Q_i\,p_i^2 \,   /\,  n_i \\
    0 & 0 & - Q_i \,  p_i^3 \,  / \, \sqrt{ m_i^2 + p_i^2 }
  \end{array}
  \right) \: .
\end{equation}
\end{linenomath}
Finally, taking the vertex invariant mass as the norm of the total
four-vector,
$m = \sqrt{k_{0,\text{tot}}^2 - |\vec k_\text{tot}|}$,
one obtains for the derivative matrix Eq.~\ref{equ:matrixHM} of the mass
constraint
\begin{linenomath}
\begin{equation}
  H_M \; = \; \left( -k_{x,\text{tot}} / m,-k_{y,\text{tot}} / m,
  -k_{z,\text{tot}}/m, k_{0,\text{tot}} / m \right) \: .
\end{equation}
\end{linenomath}



\bibliographystyle{LHCb}
\bibliography{main}


\end{document}